# Model of a Data Mining System for Personalized Therapy of Speech Disorders

*Mirela Danubianu[1], Stefan Gheorghe Pentiuc[2], Iolanda Tobolcea[3], Tiberiu Socaciu[4]*

[1] „ tefan cel Mare" University of Suceava, Faculty of Electrical Engineering and Computer Science
1 Universitatii str. 720229, Suceava, Romania
Phone +40 (0)230 524801, email:mdanub@eed.usv.ro
Web: www.usv.ro/~mdanub

[2] „ tefan cel Mare" University of Suceava, Faculty of Electrical Engineering and Computer Science
1 Universitatii str. 720229, Suceava, Romania
Phone +40 (0)230 524801, email:pentiuc@eed.usv.ro
Web: www.usv.ro/~pentiuc

[3] „A.I.Cuza" University of Iasi, Faculty of Psyhology and Educational Sciences
3 TomaCozma str. 700554, Iasi, Romania
Phone +40 (0)232 201028, email:itobolcea@yahoo.com

[4] „ tefan cel Mare" University of Suceava, Faculty of Economic Science and Public Administration
1 Universitatii str. 720229, Suceava, Romania
Phone +40 (0)230 524801, email:socaciu@seap.usv.ro

*Abstract:* **Lately, the children with speech disorder have more and more become object of specialists' attention and investment in speech disorder therapy are increasing  The development and use of information technology in order to assist and follow speech disorder therapy allowed researchers to collect a considerable volume of data. The aim of this paper is to present a data mining system designed to be associated with TERAPERS system in order to provide information based on which one could improve the process of personalized therapy of speech disorders.**

## I. INTRODUCTION

Many studies have concluded that there are children who initially may present diffuse complex problems. These problems can evolve, in time, to specific difficulties of language or speech.. The difficulty for researchers and therapists is to identify those children who have disorders that show a wide range of problems that will not resolve spontaneously or which may lead to further significant deficiencies. Based on research on patterns of development of speech and language, these children could be identified so the therapy should be conducted more efficiently.

The aim of this paper is to  presents a model for the Logo-DM system. This is a data mining system designed to be associated with TERAPERS system in order to use the data from TERAPERS for analysis and to provide new information based on which one could improve the process of therapy

## II. RELATED WORKS

We have found a lot of researches in the area of speech analyze or speech impairment therapy. Firstly there are applications regarding voice recognition [1][2][3][4] which are useful in areas such Health Care, Legal, Corporate, Education, Government, and Assistive Technology. Also, there are a lot of projects and products regarding speech recognition (e.g. Windows Speech Recognition). Other category groups the research regarding speech segmentation and the speech analyze [5][6][7]. In the area of logopaedic therapy the priorities on the international level are represented by the developing information systems that will allow the elaboration of personalized therapeutically paths. Some considered directions are: development of expert systems that personalizing the therapeutically guides to the child's evolution and the evaluation of the motivation and progresses that the child's achieves [8][9][10][11].

At the national level, little research has been conducted on the therapy of speech impairments, out of which mostly is focused on traditional areas such as voice recognition, voice synthesis and voice authentication.

Recently, TERAPERS project developed within the Center for Computer Research in the University "Stefan cel Mare" of Suceava [19] has proposed to develop a system which is able to assist teachers in their speech therapy of dislalya and to follow how the patients respond to various personalized therapy programs.

On the international level there are some data mining applications developed in areas related with logopaedic therapy, such as Health Care or Psychology [12][13][14] [15]. There are also researches regarding the possibilities of application of data mining techniques in order to classify children with speech sound disorders empirically, using factor analytic techniques [2], or to discover hidden time patterns in behavior.

The novelty of our project consist in using data mining to make predictions regarding the outcome of personalized therapy of speech disorders and to enrich the knowledge base of the expert system designed to assist the therapy in TERAPERS project[19].

### III. DATA MINING METHODS AND TECHNIQUES

Data mining is defined as the process of extracting interesting and previously unknown information from data, and it is widely accepted to be a single phase in a complex process known as Knowledge Discovery in Databases (KDD).

According to CRISP-DM [16], the reference model for this process, KDD consists of following phases:
- *business understanding*. This phase focuses on understanding the project objectives and requirements from a business perspective, on assessing of situation and determining the data mining goals.
- *data understanding* . Consist of initial data collection, describing data and verifying data quality.
- *data preparation*. Contains all activities needed to build the final data set from the initial raw data. Tasks include table, record and attribute selection and transformation and cleaning data for modeling tools.
- *modeling* Is the phase when various modeling techniques are selected and applied and the model is build.
- *evaluation* Once build, the model is tested to be certain the proper model to achieve the project objectives. At the end of this phase a decision regarding the use of the data mining results should be reached.
- *deployment* Often the created model need to be presented in a way that the customer can use it. Depending on the requirements, the deployment phase can be from simple generating of a report to complex implementing of a repeatable data mining process across the enterprise

In order to ensure that the extracted information generated by the data mining algorithms is useful, additional activities are required, like incorporating appropriate prior knowledge and proper interpretation of the data mining results.

Results obtained through the application of data mining algorithms can provide answers to two broad categories of problems: prediction and description. Using appropriate methods can solve each of these problems [17]. Figure 1 presents a summary of data mining methods.

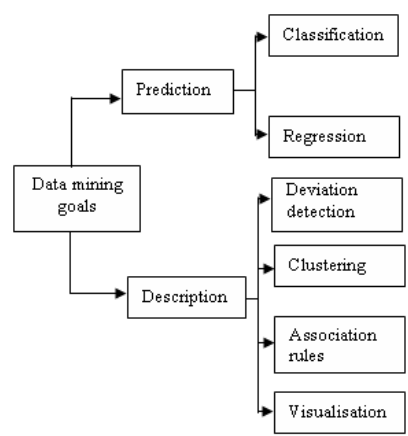

Figure 1. Summary of data mining methods

*Data classification* is a two step process. In the first step, a model is built describing a predetermined set of data classes or concepts. The model is constructed by analyzing database tuples described by attributes. Each tuple is assumed to belong to a predefined class, as determined by one of the attributes, called the class label attribute. In the context of classification, data tuples are also referred to as samples, examples, or objects. The data tuples analyzed to build the model collectively form the training data set. The individual tuples making up the training set are referred to as training samples and are randomly selected from the sample population. Since the class label of each training sample is provided, this step is also known as supervised learning (i.e., the learning of the model is 'supervised' in that it is told to which class each training sample belongs). It contrasts with unsupervised learning (or clustering), in which the class labels of the training samples are not known, and the number or set of classes to be learned may not be known in advance.

*Deviation detection* resides in detecting changes from the norm and it has many similarities with statistical analysis.

*Cluster analysis* is a set of methodologies for automatic classification of samples into a number of groups using a measure of association, so that the samples in one group are similar and samples belonging to different groups are not similar. The input for a system of cluster analysis is a set of samples and a measure of similarity (or dissimilarity) between two samples. The output from cluster analysis is a number of groups (clusters) that form a partition, or a structure of partitions, of the data set. One additional result of cluster analysis is a generalized description of every cluster, and this is especially important for a deeper analysis of the data set's characteristics.

*Association rule mining* finds interesting association or correlation relationships among a large set of data items. A typical example of association rule mining is market basket analysis. This process analyzes customer buying habits by finding associations between the different items that customers place in their shopping baskets".

### IV. MAY BE APPLIED DATA MINING TECHNIQUES IN THE SPEECH DISORDERS AREA?

In the area of speech therapy on can use the following data mining tasks[18]:
- *classification* which aims to put the persons with speech disorders in some predefined segments. This method allows estimating the dimensions and the structures of the different groups. Classification uses the information contained in the set of predictor variables (e.g. those relating to the personal or family anamnesis, or to the lifestyle), for relating the persons with the different segments.
- *clustering* which is able to forms groups of persons with speech disorders based on the similarity of some characteristics, but it does not use the apriori defined groups. It is an important task since it helps the speech therapists to understand who their patients are. For instance by clustering they can discover a subgroup of a predetermined segment who has an homogeneous behavior related to different therapy methods who can be efficient targeted by a specific therapy.
- *association rules* which find the connections between data. An important task of this method should be to determine why a certain therapy programme has had success for a segment of patients while for another segment it was inefficient.

On may draw the conclusion that data mining can be a useful tool in analyze and the design of the speech therapy. But it is necessary to point out the follow limit:

The data mining techniques analyze only the data collected from the persons who have finish the therapy program. The applications of data mining generate information from the patterns obtained from the programs system corresponding to the assistance and tracking of the speech therapy. These patterns can help to predict the evolution of the children that are in the therapy process or to the planning of an adequate schema of therapy for those.

Data mining cannot supply information about speech impairments, persons or behaviors that are not in the analyzed database.

### V. LOGO-DM SYSTEM. OBJECTIVES

The idea of trying to improve the quality of logopaedic therapy by applying some data mining techniques started from TERAPERS project developed within the Center for Computer Research in the University "Stefan cel Mare" of Suceava [19]. This project has proposed to develop a system which is able to assist teachers in their speech therapy of dislalya and to follow how the patients respond to various personalized therapy programs.

This system contains two components: an intelligent system installed on each speech therapist's office computer, called LOGOMON, and a mobile system used as a friend of child therapy. The two systems are connected. This structure is presented in Figure 1.

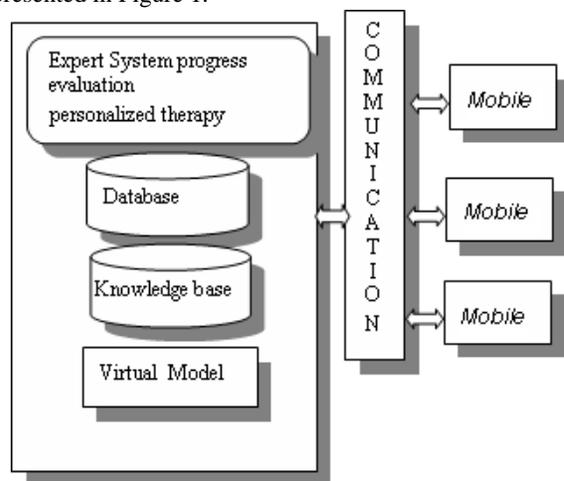

Figure 1. TERAPERS System

The intelligent system is the fix component of the system and it is installed on each speech therapist's office computer. This system includes the following parts:
- an evaluation module of children's progress
- an expert system that will produce inferences based on the data presented by the evaluation module.
- a virtual module of the mouth, that would allow the presentation of every hidden move that occur in speaking.

The mobile device of personalized therapy has two main objectives. It is used by the child in order to resolve the homework prescribed by the speech therapist and delivers to the intelligent system a personalized activity report of the child.

Due to the needs of efficient use of time or due to the economic needs, now there are of interest information such as[20]:
- how is the estimated duration of therapy for a particular case,

- what is the predicted final state for a child or what will be its state at the end of various stages of therapy,
- what are the best exercises for each case and how can dose their effort for effectively solve these exercises,
- how is associated the family receptivity - which is an important factor in success of the therapy - with other aspects of family and personal anamnesis.

All this may be the subject of predictions obtained by applying data mining techniques on data collected by using TERAPERS. It is also interesting, as part of the knowledge discovered by data mining algorithms, to be used to enrich the knowledge base of expert system embedded in LOGOMON. To achieve this goal we propose the development of Logo-DM system.

Data collected by the LOGOMON system together with data from other sources (eg demographic data, medical or psychological research) is the set of raw data that will be the subject of data mining.

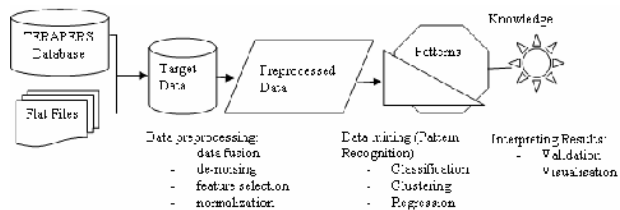

Figure 2. LOGO-DM view of the end to end data mining process

Figure 2 [20] presents the complete sequence of operations applied on these data.

From the beginning we must remark that we have a relational database, characterized by a high degree of normalization, making the various characteristics to be in different tables. Another remark concerning the data contained in the database relates to the encoding characteristics by the values specified in the various fields. These codes were converted so that the data set on which apply data mining algorithms contain descriptive understandable values.

For example, the table „*Fise*" containing data on personal and family anamnesis of children has a field concerning health problems which is of type numeric, having the following values whose meanings are:
- 10000 - serious illness
- 1000 – psychological trauma
- 100 – surgery
- 10 – accidents
- 1 – other problems

It is preferable to change the numerical values during the data preprocessing step, so that the final data set contains the descriptive values of characteristics.

We found that, due to the fact that in LOGOMON database schema some fields values are not restricted to *not null*, these values are partially filled. They should be filled automatically or if this is not possible manually. It is a problem that can be solved in time by setting those fields that have relevance for analysis and by configuring them as *not null* in data sources tables.

Creating target data set is accomplished through a join of tables containing useful features followed by a projection on a superset of appropriate attributes, as is shown in (1)

$$\prod_{I_i}(T_1 \bowtie T_2 \bowtie ... \bowtie T_k) \quad (1)$$

where: $I_i$ is a superset of the attributes regarding the useful characteristics

$T_1 ... T_k$ is the set of tables containing the useful attributes. All of these attributes must be in the projection list

As most data mining techniques were not designed to cope with large amounts of irrelevant features, combining them with feature selection techniques has become a necessity in many applications [19][20]. The most important objectives of feature selection are: to avoid over fitting and improve model performance, i.e. prediction performance in the case of supervised classification and better cluster detection in the case of clustering; to provide faster and more cost-effective models and to gain a deeper insight into the underlying processes that generated the data. In the feature selection problem, we are given a fixed set of candidate features for use in a learning problem, and must select a subset that will be used to train a model that is "as good as possible" according to some criterion.

Firstly we remark that the data from the set obtained by previous operations from LOGOMON database contains discrete and nominal values.

So, for classification we used as filter metrics mutual information. Although in mathematic sense it is not a true metric or distance measure it may be regarded as a score calculated between a set of candidate features and the category wanted to exit.

We have used for feature selection a variant of the mRMR method [21] for categorical values. The criterion used is related to minimizing redundancy and maximizing relevance to the chosen characteristics. The result of tests performed on data collected from LOGOMON and prepared as mentioned above, revealed that for classification, the minimum error is obtained if we deal with a number between 50 and 55 features selected.

The next step consists in applying different data mining algorithms on the data previously obtained. Starting from the needs of information and from the analysis of available data in the database, we can make the following observations.

According with the nature of the predictor data is suitable to use, for classification, decision trees. An example of classification for the patients, according to their possible status at the end of therapy, is: corrected (C), improved (I) or stationary (S).

In order to mine the association rules the system uses an Apriori algorithm implementation, applied on the data processed so as to obtain a transactional structure.

It should be noted that for the moment, the volume of data on which work is relatively low due to the fact that TERAPERS, which is the main source of these data is operational only for several months.

## VI. CONCLUSIONS AND FUTURE WORK

The data mining technology may be a useful tool for the logopaedic therapy. Logo-DM system is an attempt of development of a data mining system which aims to use data from TERAPERS system in order to find patterns able to supply information that allow the implementation of optimized personalized therapy programs that are adapted to the characteristics of each person.

This may conduct to reduction of duration of therapy, to increasing the possibilities of achieving superior results and finally to lower cost of the therapy.

Since, for the moment the volume of data on which work is low, we intend to cope with therapists from Regional Speech Therapy Center of Suceava in order to collect enough data to obtain relevant results. Also we will test new implementations of data mining algorithms for finding the best one for our system.

**Mirela Danubianu** is lecturer at "Stefan cel Mare" University of Suceava, Faculty of Electrical Engineering and Computers Science, Computers Chair . She has a MS in Computer Science at University of Craiova, Faculty of Electrical Engineering (1985 – Automatizations and Computers) and other in Economics at Stefan cel Mare University of Suceava, Faculty of Economics (2009 - Management). She is PhD in Computers Science at "Stefan cel Mare" University of Suceava, Faculty of Electrical Engineering and Computers Science (2006 - Contributions to the development of data mining and knowledge methods and techniques).. Current and past works: databases theory and implementation, data mining and data warehousing, application of advanced information technology in economics and health care area.

**Stefan Gheorghe Pentiuc** is professor at "Stefan cel Mare" University of Suceava, dean of Faculty of Electrical Engineering and Computer Science. From 1993 he is Ph. D. in Computer Science and Engineering at the "Politehnica" University of Bucharest. Current and past works: pattern recognition, distributed artificial intelligence, programming on Internet, algorithms and Java technologies

**Iolanda Tobolcea** is associate professor at Faculty of Psychology and Educational Science at "Alexandru Ioan Cuza Univesity of Iasi, Department of Clinical Psychology and Special Education. From 1997 she is PhD in Psychology. Current and past works: speech and language disorders therapy, psychotherapy.

**Tiberiu Socaciu** has two MS in CS at Babes-Bolyai University of Cluj, Faculty of Mathematics (1995 - Informatics and 1996 – Design and Implementation of Complex Systems). He is PhD candidate in Informatics at Babes-Bolyai University of Cluj, Faculty of Mathematics (Computer clusters with applications in HPC) and Economic Informatics at Alexandru Ioan Cuza University of Iasi, Faculty of Economics (Parallelizations in DSSs for financial engineering). He is lecturer in Informatics at Stefan cel Mare University of Suceava, Faculty of Economics, Informatics Chair and Vasile Goldis West University of Arad, Faculty of Informatics, Informatics Chair. Current and past works: financial computational engineering, computer clusters, parallel calculus, collaborative/distributed calculus, symbolic calculus/computer algebra, open systems, networking